\documentstyle[aps, epsf, floats]{revtex}

\begin{document}
\title{Many-body Theory {\it vs} Simulations for the pseudogap in the Hubbard model}
\author{S. Moukouri, S. Allen, F. Lemay, B. Kyung, D. Poulin, Y.M. Vilk$^{1}$ and
A.-M. S. Tremblay$^{2}$}
\address{D\'{e}partement de physique and Centre de recherche sur les propri\'{e}t\'{e}%
s \'{e}lectroniques de mat\'{e}riaux avanc\'{e}s\\
$^{2}$Institut canadien de recherches avanc\'{e}es\\
Universit\'{e} de Sherbrooke, Sherbrooke, Qu\'{e}bec, Canada J1K 2R1.\\
$^{1}$2100 Valencia Dr. apt. 406, Northbrook, IL 60062}
\date{\today }
\maketitle

\begin{abstract}
The opening of a critical-fluctuation induced pseudogap (or precursor
pseudogap) in the one-particle spectral weight of the half-filled
two-dimensional Hubbard model is discussed. This pseudogap, appearing in our
Monte Carlo simulations, may be obtained from many-body techniques that use
Green functions and vertex corrections that are at the same level of
approximation. Self-consistent theories of the Eliashberg type (such as the
Fluctuation Exchange Approximation) use renormalized Green functions and
bare vertices in a context where there is no Migdal theorem. They do not
find the pseudogap, in quantitative and qualitative disagreement with
simulations, suggesting these methods are inadequate for this problem.
Differences between precursor pseudogaps and strong-coupling pseudogaps are
also discussed.
\end{abstract}


\section{Introduction}

The two-dimensional Hubbard model is one of the key paradigms of many-body
Physics and is extensively studied in the context of the cuprate
superconductors. While there is now a large consensus about the fact that at
half-filling $\left( n=1\right) $ the ground state has long-range
antiferromagnetic (or spin-density wave) order,\cite{hirsh,white1} the route
to this low-temperature phase is still a matter of controversy when the
system is in the weak to intermediate coupling regime. In this regime, we
know that the Mermin-Wagner theorem precludes a spin-density-wave phase
transition at finite temperature but the issue of whether there is, or not,
a precursor pseudogap at finite temperature in the single-particle spectral
weight $A(k_{F},\omega )$ is still unresolved. Different many-body
approaches give {\it qualitatively} different answers to this pseudogap
question. In particular, the widely used self-consistent Fluctuation
Exchange Approximation (FLEX)\cite{bickers1} does not find a pseudogap in
the $d=2$ repulsive Hubbard model for any filling. A study\cite{deisz} of
lattices of up to $L=128$ found that as the temperature is reduced the
quasiparticle peak in $A(k_{F},\omega )$ smears considerably while remaining
maximum at $\omega =0$, signaling a deviation from the Fermi liquid behavior
but no pseudogap. The same qualitative answer is found for attractive
models. By contrast, the many-body approach that has given to date the best
agreement with simulations of both static\cite{VilkChen} and imaginary-time
quantities \cite{vilk2} concludes to the existence of a precursor
single-particle pseudogap in the weak to intermediate coupling regime, for
both the attractive and repulsive $d=2$ Hubbard model, whenever the ground
state has long-range order. While we will restrict ourselves to the $d=2$
repulsive model at half-filling, our results will be relevant to the more
general question of the pseudogap since small changes in filling or changes
from repulsive to attractive case\cite{vilk1}\cite{Allen99} do not generally
necessitate fundamental changes in methodology. And the question of
many-body methodology is one of our main concerns here. Further comments on
the regime we do not address here, namely the strong-coupling regime, appear
in the concluding paragraphs.

One may think that numerical results have already resolved the pseudogap
issue defined above, but this is not so. Early Quantum Monte Carlo (QMC)
data analytically continued by the Maximum Entropy method concluded that
precursors of antiferromagnetism in $A({\bf k},\omega )$ were absent at any
non-zero temperature in the weak to intermediate coupling regime ($U<8t$, $U$
is the Coulomb repulsion term and $t$ the hopping parameter)\cite{white2}. A
subsequent study in which a singular value decomposition technique was used
instead of Maximum Entropy, concluded to the opening of a pseudogap in $%
A(k_{F},\omega )$ at low temperatures\cite{creffield}. Each of the two
techniques has limitations. The singular value decomposition can achieve a
better resolution at low frequencies, but we find that the quality of the
spectra is influenced by the profile function introduced to limit the range
of frequencies. Another difficulty is that it leads to negative values of $A(%
{\bf k},\omega )$. As far as Maximum Entropy is concerned, recent advances 
\cite{jarrell}, that we will use here, have made this method more reliable
than the Classic version applied in Ref.\cite{white2}.

In this paper, we address the issue of the pseudogap in the $d=2,$ $n=1$
Hubbard model at weak to intermediate coupling, but it will be clear that
the general conclusions are more widely applicable. We present QMC results
and show that the finite-size behavior obtained for $A(k_{F},\omega )$ is
correctly reproduced by the method of Ref.\cite{vilk1}. We also introduce a
slight modification of the latter approach that makes the agreement even
more quantitative. This many-body approach allows us to extrapolate to
infinite size and show that the pseudogap persists even in lattices whose
sizes are greater than the antiferromagnetic correlation length $\xi $,
contrary to the statements made earlier\cite{white2}. These sizes cannot be
reached by QMC when the temperature is too low. We confirm that at low
enough temperatures, the peak at $\omega =0$ at the Fermi wave vector is
replaced by a minimum, corresponding to the opening of a pseudogap\cite
{vilk1} and by two side peaks that are precursors of the Bogoliubov
quasiparticles. In contrast, we find that the $A(k_{F},\omega )$ calculated
by FLEX on small lattices are {\it qualitatively} different from those of
QMC and do not have the correct size dependence. Since all many-body
techniques involve some type of approximation, their reliability should be
gauged by their capacity to reproduce, at least qualitatively, the Monte
Carlo results in regimes where the latter are free from ambiguities. We thus
conclude that Eliashberg-type approaches such as FLEX are unreliable in the
absence of a Migdal theorem and that there is indeed a pseudogap in the weak
to intermediate coupling regime at half-filling. It is likely, but not yet
unambiguously proven, that consistency between the Green functions and
vertices used in the many-body calculation is crucial to obtain the
pseudogap.

\section{Many-body approach}

\begin{figure}%
%
\centerline{\epsfxsize 8cm \epsffile{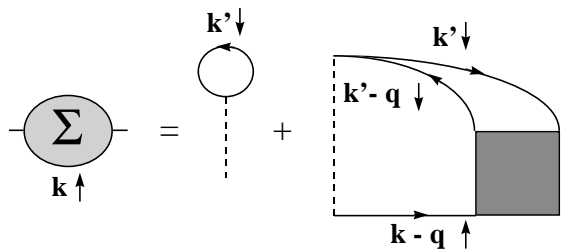}}%
%
\caption{Formally exact diagrammatic representation of the self-energy
in the Hubbard model. The square is the fully reducible four-point vertex}%
%
\label{fig1}%
%
\end{figure}%
%
Many-body techniques of the paramagnon type\cite{Berk} do lead to a
pseudogap but they usually have low-temperature problems because they do not
satisfy the Mermin-Wagner theorem. No such difficulty arises in the approach
of Ref.\cite{vilk1}. This method proceeds in two stages. In the zeroth order
step, the self-energy is obtained by a Hartree-Fock-type factorization of
the four-point function with the {\it additional constraint }that the
factorization is exact when all space-time coordinates coincide.\cite
{NoteAddConst} Functional differentiation, as in the Baym-Kadanoff approach 
\cite{Baym}, then leads to a momentum- and frequency-independent irreducible
particle-hole vertex for the spin channel that satisfies\cite{VilkChen} $%
U_{sp}=U\left\langle n_{\uparrow }n_{\downarrow }\right\rangle /\left(
\left\langle n_{\uparrow }\right\rangle \left\langle n_{\downarrow
}\right\rangle \right) $. The irreducible vertex for the charge channel is
too complicated to be computed exactly, so it is assumed to be constant and
its value is found by requiring that the Pauli principle in the form $%
\left\langle n_{\sigma }^{2}\right\rangle =\left\langle n_{\sigma
}\right\rangle $ be satisfied. More specifically, the spin and charge
susceptibilities now take the form $\chi _{sp}^{-1}\left( q\right) =\chi
_{0}(q)^{-1}-\frac{U_{sp}}{2}$ and $\chi _{ch}^{-1}\left( q\right) =$ $\chi
_{0}(q)^{-1}+\frac{U_{ch}}{2}$ with $\chi _{0}$ computed with the Green
function $G_{\sigma }^{0}$ that contains the self-energy whose functional
differentiation gave the vertices. This self-energy is constant,
corresponding to the Hartree-Fock-type factorization.\cite{NoteSelfNotHF}
The susceptibilities thus satisfy conservations laws,\cite{Baym} the
Mermin-Wagner theorem, as well as the Pauli principle $\left\langle
n_{\sigma }^{2}\right\rangle =\left\langle n_{\sigma }\right\rangle $
implicit in the following two sum rules 
\begin{eqnarray}
\frac{T}{N}\sum_{q}\chi _{sp}\left( q\right) &=&\left\langle \left(
n_{\uparrow }-n_{\downarrow }\right) ^{2}\right\rangle =n-2\left\langle
n_{\uparrow }n_{\downarrow }\right\rangle  \label{Suscep} \\
\frac{T}{N}\sum_{q}\chi _{ch}\left( q\right) &=&\left\langle \left(
n_{\uparrow }+n_{\downarrow }\right) ^{2}\right\rangle
-n^{2}=n+2\left\langle n_{\uparrow }n_{\downarrow }\right\rangle -n^{2} 
\nonumber
\end{eqnarray}
where $n$ is the density. We use the notation, $q=\left( {\bf q,}%
iq_{n}\right) $ and $k=\left( {\bf k,}ik_{n}\right) $ with $iq_{n}$ and $%
ik_{n}$ respectively bosonic and fermionic Matsubara frequencies. We work in
units where $k_{B}=1,$ $\hbar =1,$ lattice spacing and hopping $t$ are unity$%
.$ The above equations, in addition to\cite{VilkChen} $U_{sp}=U\left\langle
n_{\uparrow }n_{\downarrow }\right\rangle /\left( \left\langle n_{\uparrow
}\right\rangle \left\langle n_{\downarrow }\right\rangle \right) $, suffice
to determine the constant vertices $U_{sp}$ and $U_{ch}$. This Two-Particle
Self-Consistent approach will be used throughout this paper, unless we refer
to FLEX calculations.

Once the two-particle quantities have been found as above, the next step of
the approach of Ref.\cite{vilk1}, consists in improving the approximation
for the single-particle self-energy by starting from an exact expression
where the high-frequency Hartree-Fock behavior is explicitly factored out.
One then substitutes in the exact expression the irreducible low frequency
vertices $U_{sp}$ and $U_{ch}$ as well as $G_{\sigma }^{0}(k+q)$ and $\chi
_{sp}(q),\chi _{ch}(q)$ computed above. In the original approach\cite{vilk2}
the final formula reads 
\begin{equation}
\Sigma _{\sigma }^{\left( \ell \right) }(k)=Un_{-\sigma }+\frac{U}{4}\frac{T%
}{N}\sum_{q}\left[ U_{sp}\chi _{sp}(q)+U_{ch}\chi _{ch}(q)\right] G_{\sigma
}^{0}(k+q).  \label{Self-long}
\end{equation}
Irreducible vertices, Green functions and susceptibilities appearing on the
right-hand side of this expression are all at the same level of
approximation. They are the same as those used in the calculations of Eq.(%
\ref{Suscep}), hence they are consistent in the sense of conserving
approximations. The resulting self-energy $\Sigma _{\sigma }^{\left( \ell
\right) }(k)$ on the left hand-side though is at the next level of
approximation so it differs from the self-energy entering the right-hand
side.

There is, however, an ambiguity in obtaining the self-energy formula Eq.(\ref
{Self-long}). Within the assumption that only $U_{sp}$ and $U_{ch}$ enter as
irreducible particle-hole vertices, the self-energy expression in the
transverse spin fluctuation channel is different. To resolve this paradox,
consider the exact formula for the self-energy represented symbolically by
the diagram of Fig.1. In this figure, the square is the fully reducible
vertex $\Gamma \left( q,k-k^{\prime },k+k^{\prime }-q\right) .$ In all the
above formulas, the dependence of $\Gamma $ on $k+k^{\prime }-q$ is
neglected since the particle-particle channel is not singular. The
longitudinal version of the self-energy Eq.(\ref{Self-long}) takes good care
of the singularity of $\Gamma $ when its first argument $q$ is near $\left(
\pi ,\pi \right) .$ The transverse version does the same for the dependence
on the second argument $k-k^{\prime }$, which corresponds to the other
particle-hole channel. One then expects that averaging the two possibilities
gives a better approximation for $\Gamma $ since it preserves crossing
symmetry in the two particle-hole channels. Furthermore, one can verify that
the longitudinal spin fluctuations in Eq.(\ref{Self-long}) contribute an
amount $U\left\langle n_{\uparrow }n_{\downarrow }\right\rangle /2$ to the
consistency condition\cite{vilk2} $\frac{1}{2}{\rm Tr}\left( \Sigma ^{\left(
\ell \right) }G^{0}\right) =$ $U\left\langle n_{\uparrow }n_{\downarrow
}\right\rangle $ and that each of the two transverse spin components also
contribute $U\left\langle n_{\uparrow }n_{\downarrow }\right\rangle /2$ to $%
\frac{1}{2}{\rm Tr}\left( \Sigma ^{\left( t\right) }G^{0}\right) =$ $%
U\left\langle n_{\uparrow }n_{\downarrow }\right\rangle .$ Hence, averaging
Eq.(\ref{Self-long}) and the expression in the transverse channel also
preserves rotational invariance. In addition, one verifies numerically that
the exact sum rule\cite{vilk1} $-\int d\omega ^{\prime }%
\mathop{\rm Im}%
\left[ \Sigma _{\sigma }\left( {\bf k,}\omega ^{\prime }\right) \right] /\pi
=U^{2}n_{-\sigma }\left( 1-n_{-\sigma }\right) $ determining the
high-frequency behavior is satisfied to a higher degree of accuracy. As a
consistency check, one may also verify that $\frac{1}{2}{\rm Tr}\left(
\Sigma ^{\left( t\right) }G^{\left( t\right) }\right) $ differs by only a
few percent from$\frac{1}{2}{\rm Tr}\left( \Sigma ^{\left( t\right)
}G^{0}\right) .$ We will thus use a self-energy formula that we call
``symmetric'' 
\begin{equation}
\Sigma _{\sigma }^{\left( s\right) }(k)=Un_{-\sigma }+\frac{U}{8}\frac{T}{N}%
\sum_{q}\left[ 3U_{sp}\chi _{sp}(q)+U_{ch}\chi _{ch}(q)\right] G_{\sigma
}^{0}(k+q).  \label{Self}
\end{equation}
$\Sigma _{\sigma }^{\left( s\right) }(k)$ is different from so-called
Berk-Schrieffer type expressions\cite{Berk} that do not satisfy\cite{vilk1}
the consistency condition between one- and two-particle properties, $\frac{1%
}{2}{\rm Tr}\left( \Sigma G\right) =$ $U\left\langle n_{\uparrow
}n_{\downarrow }\right\rangle .$

\begin{figure}%
%
\centerline{\epsfxsize 12cm \epsffile{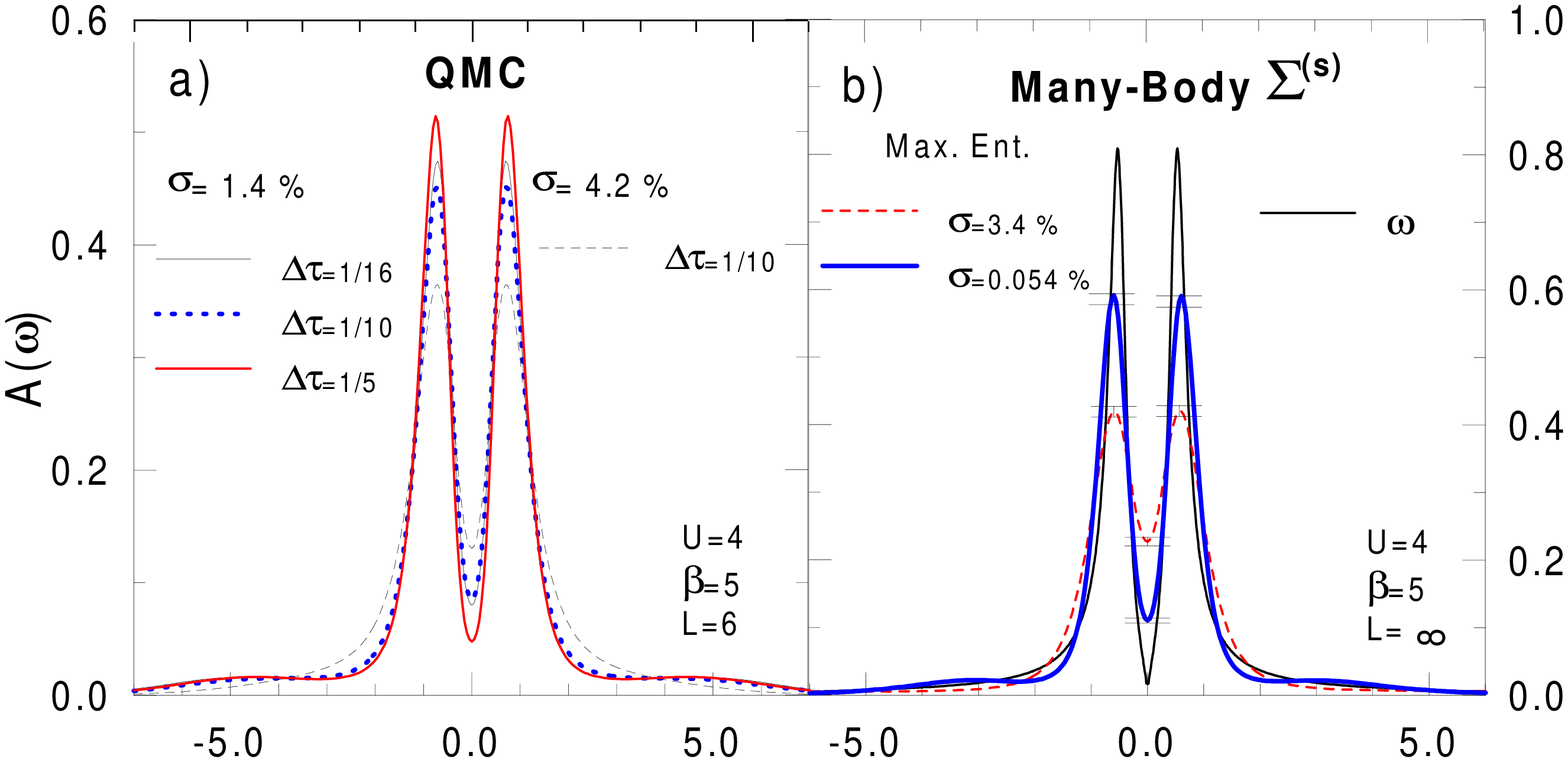}}%
%
\caption{For $U=4$, $\beta=5$, $n=1$, ${\bf k}=(0,\pi)$, 
effect of various other calculational parameters: (a) QMC for $L=6$.  
Thick dotted line for $\Delta \tau=1/10$ and $\sigma=1.4\%$.
The latter is the average of the error on $G(\tau)$ 
normalized by $G(\tau)$ itself. Calculations with the same 
$\sigma$ but for $\Delta\tau=1/5$ and $1/16$ 
are also shown. 
Thin dashed line is for $\Delta\tau=1/10$ but $\sigma=4.2\%$ on
$G(\tau)$. (b) Thin solid line, real-frequency calculation using Eq.(\ref{Self}),
for an infinite system. 
Also shown, Maximum Entropy inversion of $G(\tau)$ with same $\sigma$
as in Fig.3 and a smaller one.}%
%
\label{fig2}%
%
\end{figure}%
%
In comparing the above self-energy formulas with FLEX, it is important to
note that the same renormalized vertices and Green function appear in both
the conserving susceptibilities and in the self-energy formula Eq.(\ref{Self}%
). In the latter, one of the external vertices is the bare $U$ while the
other is dressed ($U_{sp}$ or $U_{ch}$ depending on the type of fluctuation
exchanged). This means that the fact that Migdal's theorem does not apply
here is taken into account. This technique is to be contrasted with the FLEX
approximation where all the vertices are bare ones, as if there was a Migdal
theorem, while the dressed Green functions appear in the calculation. The
irreducible vertex that is consistent with the dressed Green function is
frequency and momentum dependent, contrary to the bare vertex appearing in
the FLEX self-energy expression. In this Eliashberg-type self-consistent
approach then, the Green functions are treated at a high level of
approximation while all the vertices are bare, zeroth order ones. In other
words, the basic elements of the perturbation theory are treated at
extremely different levels of approximation.

\section{Monte Carlo {\it vs} many-body calculations}

Our Monte Carlo results were obtained with the determinantal method\cite
{white1} using typically $10^{5}$ Monte Carlo updates per space-time point.
The inverse temperature is $\beta =5$, the interaction strength is $U=4$ and
periodic boundary conditions on a square lattice are used. Other details
about the simulations may be found in the captions. Our detailed analysis is
for the single-particle spectral weight $A\left( {\bf k,}\omega \right) $ at
the wave-vector $k=\left( 0,\pi \right) $ but other wave vectors will also
be shown in the last figure of the paper. The Monte Carlo results are
influenced by the statistical uncertainty, by the systematic error
introduced through imaginary-time discretization, $\Delta \tau $, and by the
finite size, $L$, of the system. The two calculations with $\Delta \tau
=1/10 $ in Fig.2a show that increasing the number of QMC sweeps (smaller $%
\sigma $, defined in Fig.2) leads to a more pronounced pseudogap. The same
figure also shows calculations with the same $\sigma $ but different values
of $\Delta \tau $ (systematic error is of order $\left( \Delta \tau \right)
^{2}$). For $\Delta \tau \sim 1/10$, the decrease in pseudogap depth with
decreasing $\Delta \tau $ becomes less than the accuracy achievable by the
maximum entropy inversion. If the pseudogap persists when $L\rightarrow
\infty $ at fixed $\sigma $ and fixed $\Delta \tau =1/10$ it should be even
more pronounced with a larger number of QMC sweeps (smaller $\sigma $). The
size analysis needs to be done however in more detail since increasing the
system size $L$ at fixed $\sigma $ and $\Delta \tau $ leads to a smaller
pseudogap, as shown on the top left panel of Fig.3a.

\begin{figure}%
%
\centerline{\epsfxsize 12cm \epsffile{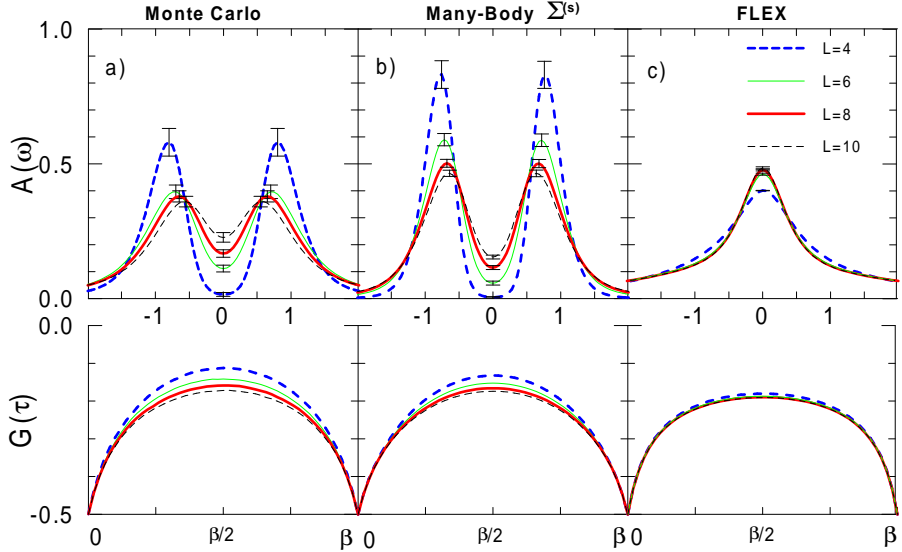}}%
%
\caption{Size dependent results for various types of calculations 
for $U=4$, $\beta=5$, $n=1$, ${\bf k}=(0,\pi)$, $L=4,6,8,10$ and
average relative errors $\sigma=3.4\%$ on $G(\tau)$. Upper panels show
$A({\bf k},\omega)$ extracted from Maximum Entropy on $G(\tau)$ shown on
the corresponding lower panels. Each $G(\tau)$ has $50$ points. (a) QMC. 
(b) Many-body using Eq.(\protect\ref{Self}). (c) FLEX.}%
%
\label{fig3}%
%
\end{figure}%
%
It is customary to analytically continue imaginary time QMC using the
Maximum Entropy algorithm\cite{jarrell}. To provide a faithful comparison
with the many-body approaches, we use the imaginary-time formalism for these
methods and analytically continue them for the same number of imaginary-time
points, using precisely the same Maximum entropy approach as for QMC. While
the round off errors in the many-body approaches are very small, it is
preferable to artificially set them equal to those in the corresponding QMC
simulations to have the same degree of smoothing. Many-body results from the
symmetric self-energy formula $\Sigma ^{\left( s\right) },$ Eq.(\ref{Self}),
for an infinite system are shown in Fig.2b. The thin solid line is a direct
real-frequency calculation in the infinite-size limit. Maximum Entropy
inversions of the $L\rightarrow \infty $ value of the many-body $G\left(
\tau \right) $ shown on the same figure illustrate that with increasing
accuracy the real-frequency result is more closely approximated. This
confirms that Maximum Entropy simply smooths the results when artificially
large errors are introduced in the analytical results.\cite{Note1} For this
parameter range, the effects are appreciable but do not change qualitatively
the results. Even the widths of the peaks are not too badly reproduced by
Maximum Entropy. The error bars are obtained from the Maximum-Entropy
Bayesian probability for different regularization parameters $\alpha .$\cite
{jarrell} They are clearly a lower bound.

In Fig.3, we show the spectra obtained for three techniques for system sizes 
$L=4,$ $6,$ $8$ and $10$. The left-hand panel is the QMC data, the middle
panel is obtained from $\Sigma ^{\left( s\right) }$ Eq.(\ref{Self}) while
the last panel is for FLEX. The latter results for much larger lattices are
not much different from those for the $8\times 8$ system. Since $G\left( 
{\bf k,}\tau \right) =-\int \frac{d\omega }{2\pi }\frac{e^{-\omega \tau }}{%
e^{-\beta \omega }+1}A\left( {\bf k},\omega \right) ,$ the nearly flat ($%
\tau $-independent) portion in $G\left( {\bf k,}\tau \right) $ of the lower
right-hand panel leads, in FLEX, to a maximum in $A\left( k,\omega \right) $
at $\omega =0,$ contrary to the Monte Carlo results. By contrast, as can be
seen by comparing the middle and left panels, the agreement between Eq.(\ref
{Self}) and QMC is very good, except for the height of the peaks. The
finite-size dependence of the pseudogap for both QMC and Eq.(\ref{Self}) is
similar: as the size increases, the depth of the pseudogap decreases. Some
of the finite-size effects are present in the vertices $U_{sp}$ and $U_{ch}$.

Fig.4a compares three results for the $L=6$ system: QMC (thick solid line),
and the many-body approach of Ref.\cite{vilk1} using either the symmetric $%
\Sigma ^{\left( s\right) }$ (Eq.(\ref{Self}), dotted line) or the
longitudinal $\Sigma ^{\left( \ell \right) }$ (Eq.(\ref{Self-long}), thin
solid line) self-energy formulas. In imaginary time, the agreement between
QMC and $\Sigma ^{\left( s\right) }$ is striking. The position of the peaks
in QMC also agrees better with the symmetric version $\Sigma ^{\left(
s\right) }$, Eq.(\ref{Self}).

For the lattice sizes where the Monte Carlo data are qualitatively similar
to those of Ref.\cite{white2}, and hence uncontroversial, Fig.3 has shown
that there is a many-body approach that gives good agreement with the
simulations. Although this many-body approach is not rigorous, especially
deep in the pseudogap regime where it is mostly an extrapolation method\cite
{vilk1}, these tests suggest that it can give an understanding of
finite-size effects in QMC data. There are two intrinsic lengths that are
relevant, namely $\xi $ the antiferromagnetic correlation length, and $\xi
_{th}$ the single-particle thermal de Broglie wavelength defined by $%
v_{F}/T. $ In simulations, $\xi $ may be estimated from the momentum-space
width of the spin structure factor and $\xi _{th}$ from the Fermi velocity
estimated from the maxima of $A\left( {\bf k},\omega \right) $ at different
wave vectors$.$ For $\beta =5,$ and $L=10$ we have $\xi \sim 3.$ At the $%
\left( \pi ,0\right) $ point, $\xi _{th}$ essentially vanishes since we are
at the van Hove singularity, hence the condition $L>\xi _{th}$ is satisfied.
If we had $\xi _{th}>L,$ one would be effectively probing the finite-size
zero-temperature quantum regime. When the condition $L>\xi _{th}$ is
satisfied, as is the case here, one has access to the finite temperature
effects we are looking for. Once agreement on the pseudogap in QMC and the
analytical approach has been established up to the regime $\xi _{th}<L<\xi ,$
the analytical approach \cite{vilk1} can be used to reach larger lattice
sizes $\left( \text{such that }\xi _{th}<\xi <L\right) $ with relatively
modest computer effort. In Fig.4b 
\begin{figure}%
%
\centerline{\epsfxsize 12cm \epsffile{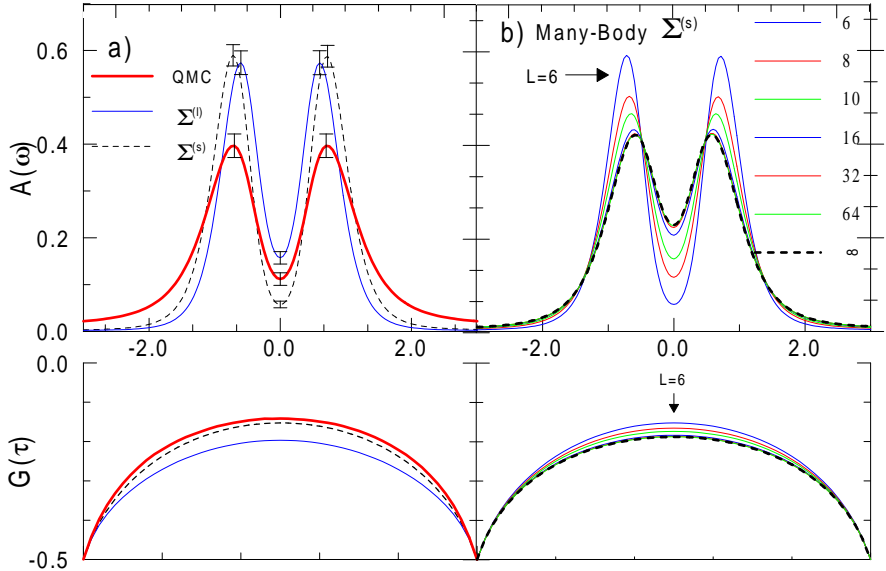}}%
%
\caption{ $U=4$, $\beta=5$, $n=1$, ${\bf k}=(0,\pi)$, and
$\sigma=3.2\%$ in Maximum Entropy.
(a) For $L=6$, thick solid line for QMC, and Many Body using 
two different self-energy formulas: dashed line for symmetric Eq.(\protect\ref{Self}), 
and thin solid line
for longitudinal Eq.(\protect\ref{Self-long}). (b) Size dependent results obtained
from symmetric version Eq.(\protect\ref{Self}) for $L=6,8,10,16,32,64$ and
infinite size (dashed line). The size dependence is monotonous.}%
%
\label{fig4}%
%
\end{figure}%
%
we show the spectra obtained by Eq.(\ref{Self}) for $L=6$ to $64$ and then
for $L=\infty $ (obtained from numerical integration). We see that the size
dependence of the pseudogap becomes negligible around $L=32$ and that the
pseudogap is quite sizable even though it is smaller than that in the
largest size available in QMC calculations $\left( L=10\right) .$ The size
dependence of the pseudogap is qualitatively similar when the longitudinal
form of the self-energy is used. We thus conclude that the pseudogap exists
in the thermodynamic limit, contrary to the conclusion of Ref.\cite{white2}.
The increase in QMC noise with increasing system size in the latter work may
partly explain the different conclusion.

The last figure, Fig.5, shows $A\left( {\bf k},\omega \right) $ obtained by
Maximum Entropy inversion of Monte Carlo data (left panel) of the many-body
approach Eq.(\ref{Self}) (middle panel) and of FLEX (right panel). Using the
symmetry of the lattice and particle-hole symmetry, $A\left( {\bf k},\omega
\right) =A\left( {\bf k+}\left( \pi ,\pi \right) ,-\omega \right) $, one can
deduce from this figure the results for {\it all} wave vectors of this $%
8\times 8$ lattice. The detailed agreement between Monte Carlo and the
many-body approach is surprisingly good for all wave vectors, even far from
the Fermi surface.

\section{Discussion}

There are two interrelated conclusions to our work. First, detailed analysis
of QMC results along with comparisons with many-body calculations show that
there is a pseudogap in the $n=1,d=2$ Hubbard model, contrary to results
obtained from previous Monte Carlo simulations\cite{white2} and from
self-consistent Eliashberg-type methods such as FLEX. Second, we have
reinforced the case that the many-body methodology described here is an
accurate and simple approach for studying the Hubbard model, even as we
enter the pseudogap regime. While any self-energy formula that takes the
form, $\Sigma \propto \sum_{q}\chi \left( q\right) G^{0}\left( k+q\right) $
will in general extrapolate correctly to a finite zero-temperature gap\cite
{Marcelja}, and hence show a pseudogap as long as $\chi \left( q\right) $
contains a renormalized classical regime,\cite{vilk1} all other approaches
we know of suffer from the following defects: they usually predict
unphysical phase transitions, they do not satisfy as many exact constraints
and in addition they do not give the kind of quantitative agreement with
simulations that we have exhibited in Figs.3 to 5. Reasons why the
mathematical structure of FLEX-type approaches fails to yield a pseudogap
have been discussed before.\cite{vilk1} The same arguments apply to the
pseudogap problem away from half-filling and for the attractive Hubbard
model as well.\cite{vilk1,Allen99,Kyung99,Allen} Since in the Hubbard model
there is no Migdal theorem to justify the neglect of vertex corrections, it
is likely, but unproven, that to obtain a pseudogap in FLEX-type approaches,
one would need to include vertex-correction diagrams that are at the same
level of approximation as the renormalized Green functions.

The physical origin of the pseudogap in the 2D Hubbard model has been
discussed at great length previously\cite{vilk1}: The precursors of
antiferromagnetism in $A\left( k_{F},\omega \right) $ are preformed
Bogoliubov quasiparticles that appear as a consequence of the influence of
renormalized classical fluctuations in two dimensions. They occur only in
low dimension when the characteristic spin relaxation rate is smaller than
temperature and when $\xi /\xi _{th}>1$. With perfect nesting (or in the
attractive Hubbard model) they occur for arbitrarily small $U.$ The
ground-state gap value\cite{NoteChargeGap} (and corresponding
single-particle pseudogap energy scale at finite $T$) depends on coupling in
a BCS-like fashion.

The previous results show that strong-coupling local particle-hole pairs are 
{\it not} necessary to obtain a pseudogap. Such local particle-hole pairs
are a different phenomenon. They lead to a single-particle Hubbard gap well
above the antiferromagnetically ordered state, {\it in any dimension }but
only{\it \ }when $U$ is large enough, in striking contrast with the
precursors discussed in the present paper. The Hubbard gap also can exist
without long-range order.\cite{NoteJ}

From a methodological point of view, the strong-coupling Hubbard gap is well
understood, in particular within the dynamical mean-field theory\cite{DMF}
or in strong-coupling perturbation expansion\cite{Pairault}. However, the
precursors of Bogoliubov quasiparticles discussed in the present paper are
unobservable in infinite dimension, where dynamical mean-field theory is
exact, because they are a low dimensional effect. It remains to be shown if $%
1/d$ expansions or other extensions of infinite-dimensional methods will
succeed in reproducing our results.\cite{Jarrel99}

Experimentally, one can distinguish a strong-coupling pseudogap from a
precursor pseudogap (superconducting or antiferromagnetic) as follows.
Ideally, if one has access experimentally to the critical quantity (spin or
pair fluctuations) the difference between the two phenomena is clear since
precursors occur only in the renormalized classical regime of these
fluctuations. If one has access only to $A\left( {\bf k},\omega \right) ,$
there are also characteristic signatures. The precursors are characterized
by a ``dispersion relation'' that is qualitatively similar to that in the
ordered state. (However the intensity of the peaks in $A\left( {\bf k}%
,\omega \right) $ does not have the full symmetry of the ordered state). By
contrast, a strong-coupling pseudogap does not show any signs of the
symmetry of the ordered state at high enough temperature.\cite{NoteJ} Also,
the temperature dependence of both phenomena is very different since
precursors of Bogoliubov quasiparticles disappear at sufficiently high
temperature in a manner that is strongly influenced by the Fermi velocity
because of the condition $\xi /\left( v_{F}/T\right) >1$.\cite
{vilk1,Allen,Preosti} Hence, even with isotropic interactions, the precursor
pseudogaps appear at higher temperatures on points of the Fermi surface that
have smaller Fermi velocity, even in cases when the zero temperature value
of the gap is isotropic. This has been verified by QMC calculations for the
attractive Hubbard model.\cite{Allen99} By contrast, at sufficiently strong
coupling, the Hubbard gap does not disappear even at relatively large
temperatures, despite the fact that $A\left( {\bf k},\omega \right) $ may
rearrange over frequency ranges much larger than temperature.\cite{Poulin}

The methods we have presented here apply with only slight modifications to
the attractive Hubbard model case where superconducting fluctuations\cite
{Kyung99} may induce a pseudogap\cite{Allen99,Allen} in the weak to
intermediate coupling regime relevant for the cuprates at that doping\cite
{Noted-wave}. Recent time-domain transmission spectroscopy experiments\cite
{Orenstein98} suggest that the renormalized classical regime for the
superconducting transition in high-temperature superconductors has been
observed. Concomitant peaks observed in photoemission experiments\cite
{Fedorov99} persist above the transition temperature in the normal state.
They may be precursors of superconducting Bogoliubov quasiparticles.\cite
{Allen99} At exactly half-filling on the other hand, the paramagnetic state
exhibits a strong-coupling (local particle-hole pairs) Hubbard gap.%
\begin{figure}%
%
\centerline{\epsfxsize 12cm \epsffile{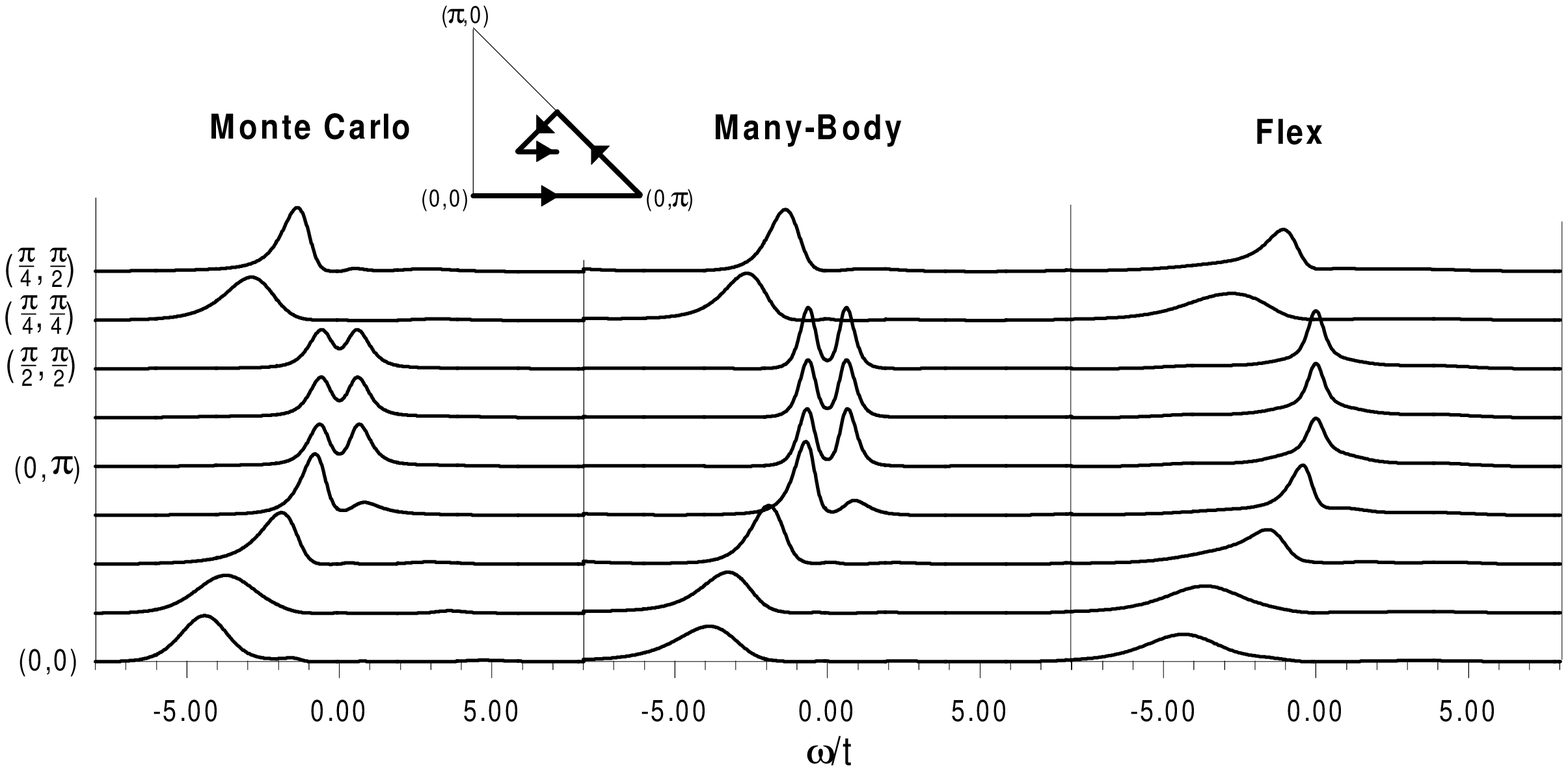}}%
%
\caption{ Single particle spectral weight $A({\bf k},\omega)$ for
$U=4$, $\beta=5$, $n=1$, and all independent wave vectors ${\bf k}$ of an
$8 \times 8$ lattice. Results obtained from
Maximum Entropy inversion of QMC data on the left panel and
many-body calculations with Eq.(\protect\ref{Self}) on the middle panel and with
FLEX on the right panel. (Relative error in all cases is about 0.3\%)}%
%
\label{fig5}%
%
\end{figure}%
%

S.M. benefited from a useful correspondence with S. R. White. We thank J.\
Deisz for extended correspondence on this subject. Contributions to the code
from H. Touchette are gratefully acknowledged. Monte Carlo simulations were
performed in part on an IBM-SP2 at the Centre d'Applications du Calcul
Parall\`{e}le de l'Universit\'{e} de Sherbrooke. This work was supported by
a grant from the Natural Sciences and Engineering Research Council (NSERC)
of Canada and the Fonds pour la formation de Chercheurs et l'Aide \`{a} la
Recherche (FCAR) of the Qu\'{e}bec government.


\begin{references}
\bibitem{NoteAddConst}  This additional constraint leads to a degree of
consistency between one- and two-particle quantities that is absent from the
standard Hartree-Fock factorization.
\bibitem{NoteJ}  However, close to the ordered state that appears at
sufficiently low temperature at large $U$ some low energy feature may
disperse with an energy scale $J=4t^{2}/U$ but its distance from zero energy
is of order $U.$
\bibitem{Noted-wave}  The model needs to be modified to obtain d-wave
pairing.
\end{references}
\end{document}